\documentclass[twocolumn,showpacs,showkeys]{revtex4}
\usepackage{graphicx}
\input{psfig.sty}

\newcommand{\bastar}{\begin{eqnarray*}}
\newcommand{\eastar}{\end{eqnarray*}}
\newskip\humongous \humongous=0pt plus 1000pt minus 1000pt

\relax
\newcommand{\be}{\begin{equation}}
\newcommand{\ee}{\end{equation}}
\newcommand{\bea}{\begin{eqnarray}}
\newcommand{\eea}{\end{eqnarray}}

\newcommand{\pro}{\partial}
\newcommand{\n}{\hat n}

\newcommand{\dfrac}{\displaystyle\frac}
\newcommand{\ba}{\begin{array}}
\newcommand{\ea}{\end{array}}

\newcommand{\nn}{\nonumber}
\newcommand{\hn}{\hat n}
\begin{document}
\title{Reinterpretation of Faddeev-Niemi Knot in Skyrme Theory}
\author{Y. M. Cho}
\email{ymcho@yongmin.snu.ac.kr}
\affiliation{School of Physics, College of Natural Sciences,
Seoul National University,
Seoul 151-742, Korea  \\
and \\
C.N. Yang Institute for Theoretical Physics, State University of
New York, Stony Brook, NY 11790, USA}

\begin{abstract}
~~~~~We identify the Faddeev-Niemi knot in Skyrme theory as a
vortex ring made of a helical baby skyrmion (a twisted
chromomagnetic vortex which is periodic in $z$-coodinate) with the
periodic ends connected together. This allows us to interpret the
knot as two quantized magnetic flux rings linked together, the
first one winding the second $m$ times and the second one winding
the first $n$ times, whose linking number $mn$ is fixed by
the Chern-Simon index of the magnetic potential.
This interpretation strongly suggests that the
Skyrme theory could also describe a very interesting low energy
physics in a completely different environment, which puts the
theory in a totally new perspective.
\end{abstract}
\maketitle

The Skyrme theory has played a very important role in physics,
in particular in nuclear physics \cite{skyr,prep,piet}.
A remarkable feature of Skyrme theory is
its rich topological structure \cite{cho01}. It has
been discovered that the theory allows not only the original
skyrmion and the baby skyrmion but also the Faddeev-Niemi knot
whose topology is fixed by $\pi_3(S^2)$ \cite{cho01,fadd1}.
Similar knots
have appeared almost everywhere recently, in atomic physics
in two-component Bose-Einstein condensates \cite{cho1,ruo,cho2},
in condensed matter physics in multi-gap superconductors \cite{cho2,baba},
in plasma physics in coronal loops \cite{fadd2},
even in high energy physics
in Weinberg-Salam model \cite{cho3}.
But at the center of all these knots
lies the Faddeev-Niemi knot of Skyrme theory \cite{cho1,cho2,cho3}.
So we need a better understanding of the knot in Skyrme theory first
to understand these knots.

The purpose of this Letter is to provide a new interpretation
of topological objects in Skyrme theory
which could allow us to construct
the Faddeev-Niemi knot in laboratories,
in particular in two-component superfluids and
two-gap superconductors.
{\it We show that the Faddeev-Niemi knot is nothing but
a vortex ring made of a helical baby skyrmion, a twisted
chromomagnetic flux which is periodic
in $z$-coordinate, with two perodic ends
connected together.
This allows us to interprete the knot as two linked
magnetic flux rings whose flux is quantized, the first one
winding the second $m$ times and the second one winding the first
$n$ times. With this we can identify
the knot quantum number
with the product of two flux quanta
$mn$, the linking number of two rings.}
We also provide an alternative interpretation of the knot,
two quantized vorticity flux rings
linked together in two-component superfluid
whose linking number becomes the knot
quantum number.

Mathematically the knot has always been described by two rings
linked together whose quantum number is given by the linking number.
In this description, however, the rings are pure mathematical 
rings defined by the preimages of the Hopf mapping 
$\pi_3(S^2)$ \cite{fadd1}. But here we show that the Faddeev-Niemi 
knot can be identified as real (i.e., physical) chromomagnetic rings linked 
together. This is a dynamical manifestation of knot, and so far 
it has never been shown that the Faddeev-Niemi knot actually has 
this dynamical manifestation. 

To understand this it is crucial to have a better understanding
of Skyrme theory, in particular of the relation among
the topological objects of the theory, first.
The Skyrme theory has a non-Abelian monopole
very similar to the Wu-Yang monopole in $SU(2)$ QCD.
And it is this monopole which plays the key role
in the theory. All the other topological objects in Skyrme theory
can be constructed
from this monopole. In particular, the baby skyrmion can be
viewed as the chromomagnetic vortex connecting
a monopole-antimonopole pair, and the Faddeev-Niemi
knot is nothing but a magnetic vortex ring made of a
twisted baby skyrmion \cite{cho01,cho02}.

To see this let $\omega$ and $\hat n$
(with ${\hat n}^2 = 1$)
be the Skyrme field and non-linear sigma field,
and let
\bea
&U = \exp (\omega \dfrac{\vec \sigma}{2i} \cdot \hat n)
= \cos \dfrac{\omega}{2} - i (\vec \sigma \cdot \hat n)
\sin \dfrac{\omega}{2}, \nn\\
&L_\mu = U\partial_\mu U^{\dagger}.
\label{su2}
\eea
With this one can write the Skyrme Lagrangian as \cite{skyr}
\bea
&{\cal L} = \dfrac{\mu^2}{4} {\rm tr} ~L_\mu^2 + \dfrac{\alpha}{32}
{\rm tr} \left( \left[ L_\mu, L_\nu \right] \right)^2,
\label{slag}
\eea
where $\mu$ and $\alpha$ are the coupling constants.
Notice that the Lagrangian has a local $U(1)$ symmetry
as well as a global $SU(2)$ symmetry.
A remarkable point of the Lagrangian is that $\omega=\pi$,
independent of $\hn$, becomes a classical solution \cite{cho01}.
So restricting $\omega$ to $\pi$, one can reduce (\ref{slag}) to
the Skyrme-Faddeev Lagrangian
\bea
{\cal L} \rightarrow -\dfrac{\mu^2}{2} (\partial_\mu
\hat n)^2-\dfrac{\alpha}{4}(\partial_\mu \hat n \times
\partial_\nu \hat n)^2,
\label{sflag}
\eea
whose equation of motion is given by
\bea
&\hn \times
\partial^2 \hn + \dfrac{\alpha}{\mu^2} ( \partial_\mu N_{\mu\nu} )
\partial_\nu \hn = 0, \nn\\
&N_{\mu\nu} = \hn \cdot (\partial_\mu \hn \times \partial_\nu \hn).
\label{sfeq}
\eea
It is this equation that allows not only
the baby skyrmion and the Faddeev-Niemi
knot but also the non-Abelian monopole.

In fact one can argue that the Lagrangian (\ref{sflag}) describes
a theory of monopole \cite{cho01,cho02}.
To see this notice that (\ref{sflag}) can be put into a very
suggestive form,
\bea
&{\cal L} = -\dfrac{\alpha}{4} \vec H_{\mu\nu}^2
- \dfrac{\mu^2} {2} \vec C_\mu^2, \nn\\
&\vec H_{\mu\nu} = \partial_\mu \vec C_\nu - \partial_\nu \vec C_\mu
+ g \vec C_\mu \times \vec C_\nu,
\label{qcdlag}
\eea
where $\vec C_\mu$ is the ``Cho connection" \cite{cho80,fadd3,gies,zucc}
\bea
\vec C_\mu = -\dfrac{1}{g} \hn \times \partial_\mu \hn.
\eea
Clearly this demonstrates that the Skyrme theory is
deeply related to QCD. Just like the $SU(2)$ QCD the Lagrangian has
the non-Abelian monopole solution \cite{cho01}
\bea
\hn = \hat r,
\label{mono}
\eea
where $\hat r$ is the unit radial vector.
Notice that the potential $\vec C_\mu$,
with (\ref{mono}), becomes nothing but the well-known
Wu-Yang monopole potential \cite{cho02,cho80}.
Moreover, the above solution becomes a solution even without the
non-linear interaction (i.e., with $\alpha=0$),
which justifies the interpretation that
the Skyrme theory is indeed a theory of monopole (interacting
with the Skyrme field $\omega$). But one has
to keep in mind that this
monopole is not an electromagnetic
monopole, but rather a non-Abelian chromomagnetic one.
Notice that
\bea
\vec H_{\mu\nu} = H_{\mu\nu} \hn = -\dfrac{1}{g}
\partial_\mu \hn \times \partial_\nu \hn
= -\dfrac{1}{g} N_{\mu\nu} \hn,
\eea
so that in this picture $N_{\mu\nu}$ becomes nothing but
the Abelian chromomagnetic field of the $U(1)$ gauge symmetry
in (\ref{sflag}).

Now, we argue that the Lagrangian (\ref{sflag}) can also
be viewed to describe a $CP^1$ model which describes a two-component
superfluid \cite{cho1,cho2}.  To see this let $\xi$ be a $CP^1$ field
which forms an $SU(2)$ doublet
and consider the $CP^1$ Lagrangian
\bea
&{\cal L} = -\dfrac {\mu^2}{2} \Big(|\pro_\mu
\xi |^2 - |\xi^\dag \pro_\mu \xi|^2 \Big) \nn\\
&- \dfrac {\alpha}{4} (\pro_\mu \xi^\dag \pro_\nu \xi
- \pro_\nu \xi^\dag \pro_\mu \xi)^2,
~~~~~(\xi^\dag \xi = 1).
\label{beclag}
\eea
But with the identification
\bea
\n = \xi^\dag \vec \sigma \xi,
\label{ndef}
\eea
we have
\bea
& (\pro_\mu \hn)^2 = 4 (|\pro_\mu \xi|^2
- |\xi^\dag \pro_\mu \xi|^2), \nn\\
&N_{\mu\nu} = \hn \cdot (\pro_\mu \hn \times \pro_\nu \hn)
= 2i (\pro_\mu \xi^\dag \pro_\nu \xi
- \pro_\nu \xi^\dag \pro_\mu \xi ) \nn\\
&= \pro_\mu C_\nu - \pro_\nu C_\mu,
\label{fmn}
\eea
where $C_\mu$ is the velocity potential of the doublet
$\xi$ \cite{cho2}
\bea
C_\mu = 2i \xi^\dag \pro_\mu \xi.
\label{cm}
\eea
This tells that the three Lagrangians (\ref{sflag}), (\ref{qcdlag}),
and (\ref{beclag}) are all identical to each other, which confirms that
the Skyrme-Faddeev theory could also be regarded as a theory of
two-component superfluid. In this view, however,
$N_{\mu\nu}$ in (\ref{sfeq}) acquires a new
meaning. It now describes
the vorticity field of the superfluid $\xi$.
It is really remarkable that the theory allows
such different interppretations.

To understand the physical meaning of the Faddeev-Niemi knot
one has to understand the helical baby skyrmion first.
To construct the desired helical vortex
we let $(\varrho,\varphi,z)$ the cylindrical coodinates,
and choose the ansatz
\bea
&\xi = \Bigg( \matrix{\cos \dfrac{f(\varrho)}{2}
\exp (-imkz-in\varphi) \cr
\sin \dfrac{f(\varrho)}{2} } \Bigg), \nn\\
&\n= \xi^\dag \vec \sigma \xi
= \Bigg(\matrix{\sin{f(\varrho)}\cos{(mkz+n\varphi)} \cr
\sin{f(\varrho)}\sin{(mkz+n\varphi)} \cr \cos{f(\varrho)}}\Bigg), \nn\\
&C_\mu = \big(\cos{f(\varrho)} +1\big)
(mk \pro_\mu z + n \pro_\mu \varphi).
\label{hvans}
\eea
With this the equation (\ref{sfeq}) is reduced to
\bea
&\Big(1+(m^2 k^2+\dfrac{n^2}{\varrho^2})
\dfrac{\sin^2{f}}{g^2 \rho^2}\Big) \ddot{f}
+ \Big( \dfrac{1}{\varrho}+ 2\dfrac{\dot{\rho}}{\rho} \nn\\
&+(m^2 k^2+\dfrac{n^2}{\varrho^2})
\dfrac{\sin{f}\cos{f}}{g^2 \rho^2} \dot{f}
+ \dfrac{1}{\varrho} (m^2 k^2-\dfrac{n^2}{\varrho^2})
\dfrac{\sin^2{f}}{g^2 \rho^2} \Big) \dot{f} \nn\\
&- (m^2 k^2+\dfrac{n^2}{\varrho^2}) \sin{f}\cos{f}=0.
\label{hveq}
\eea
So with the boundary condition
\bea
f(0)=\pi,~~f(\infty)=0,
\label{bc}
\eea
we obtain the non-Abelian vortex solutions shown in Fig.1.
There are three points that have to be
emphasized here. First, when $m=0$, the solution describes
the well-known baby skyrmion \cite{piet}. But when $m$ is not zero,
it describes a helical vortex which is periodic in $z$-coodinate.
In this case, the vortex has a non-vanishing velocity potential
(not only around the vortex but also) along the $z$-axis.
Secondly, the superfluid $\xi$ starts from
the second component at the core,
but the first component takes over completely at the infinity.
This is due to the boundary condition $f(0)=\pi$ and $f(\infty)=0$,
which assures that our solution describes a genuine
non-Abelian vortex. Thirdly, $C_\mu$ and $N_{\mu\nu}$
here can also be interpreted as the chromomagnetic potential
and field, so that one can view the helical vortex
a twisted magnetic vortex confined along the $z$-axis.
This allows us to identify
the baby skyrmion as the magnetic flux line
which connects the monopole-antimonopole pair
separated infinitely apart.

\begin{figure}
    \includegraphics[scale=0.5]{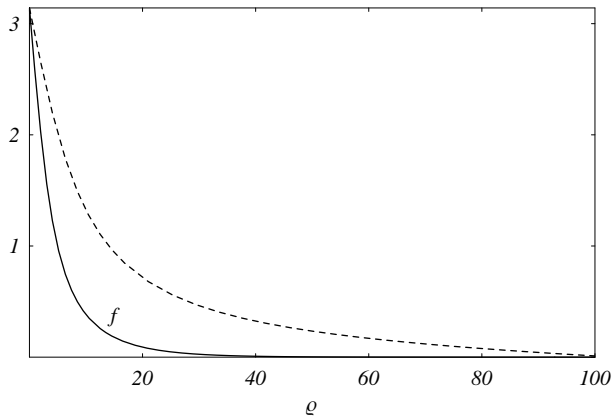}
    \caption{The non-Abelian vortex (dashed line) with $m=0,n=1$
and the helical vortex (solid line) with $m=n=1$ in two-component BEC.
Here we have put $g=\lambda=1$, $k=\rho_0/10$,
and $\varrho$ is in the unit of $1/\rho_0$.}
    \label{hbs}
\end{figure}

Remarkably the helical vortex has two distinct
chromomagnetic fluxes. To see this notice first that it has
a quantized magnetic flux along the $z$-axis,
\bea
&\phi_{\hat z} = \dfrac {}{}\int H_{\varrho\varphi} d\varrho d\varphi \nn\\
& = -\dfrac {4\pi i}{g} \int (\partial_{\varrho} \xi^{\dagger}
\partial_{\varphi} \xi - \partial_{\varphi} \xi^{\dagger}
\partial_{\varrho} \xi) d\varrho = \dfrac{4\pi n}{g}.
\label{nqn}
\eea
But due to its helical structure it also has
a quantized magnetic flux around the $z$-axis (in
one period section from $0$ to $2\pi/k$ in $z$-coordinate) given by
\bea
&\phi_{\hat \varphi} = -\dfrac {}{}\int H_{\varrho z} d\varrho dz \nn\\
&= \dfrac {4\pi i}{g} \int (\partial_{\varrho} \xi^{\dagger}
\partial_z \xi - \partial_z \xi^{\dagger}
\partial_{\varrho} \xi) \dfrac{d\varrho}{k} = \dfrac{4\pi m}{g}.
\label{mqn}
\eea
Obviously these quantized magnetic fluxes come from the
quantized magnetic potential $C_\mu$ in (\ref{hvans}), which in turn
originates from the twisted topology of the helical vortex.

The helical vortex will become
unstable unless the periodicity condition is enforced by
hand. But for our purpose it plays a very important role,
because it allows us to construct the Faddeev-Niemi
knot \cite{cho01,cho1}. To understand this, notice that we can make it
a vortex ring by smoothly connecting
two periodic ends (or by twisting the monopole-antimonopole
flux and putting the monopole and antimonopole together).
Remarkably, this vortex ring naturally
acquires the topology of a knot, and thus becomes a knot itself.
This is because
by construction this knot carries two magnetic fluxes,
$m$ unit of flux passing through the knot disk
and $n$ unit of flux passing along the knot.
Moreover the two fluxes can be thought of two unit flux rings
linked together winding each other $m$ and $n$ times,
whose linking number becomes $mn$. 

This is a dynamical manifestation of the knot. Notice that the knot 
topology has always been described by the Hopf mapping 
$\pi_3(S^2)$. When $\pi_3(S^2)$ is non-trivial the preimages 
of any two points in $S^2$ forms two rings linked together, 
whose linking number is described by 
the Chern-Simon index of the potential $C_\mu$ \cite{fadd1},
\bea
&Q = \dfrac{1}{32\pi^2} \int \epsilon_{ijk} C_i N_{jk} d^3x \nn\\
&= - \dfrac {1}{4\pi^2} \int \epsilon_{ijk} \xi^{\dagger}
\partial_i \xi ( \partial_j \xi^{\dagger}
\partial_k \xi ) d^3 x = mn.
\label{kqn}
\eea
This is the mathematical definition of the knot quantum number.
But here we have shown that this knot quantum number
is precisely the linking number of two magnetic flux rings,
which have nothing to do with the preimages of the Hopf
mapping. This tells that the knot structure is manifest even at 
the dynamical level. This point has not been well appreciated so far. 
Notice that, with the Hopf fibering of $S^3$ to
$S^2 \times S^1$, the knot quantum number
can also be viewed to represent the mapping $\pi_3(S^2)$ of $\hn$
or $\pi_3(S^3)$ of $\xi$.

Clearly the knot has a topological stability, because two
flux rings linked together can not be disjointed by a smooth
deformation of the field configuration. 
But the above analysis tells that
the topological stability is now
backed up by the dynamical stability. To see this,
notice that the quantized chromomagnetic flux of the rings
can be thought to come from the
chromoelectric supercurrent
\bea
j_\mu = \dfrac{1}{4\pi} \pro_\nu H_{\mu\nu}
= \dfrac{1}{4\pi g} (\pro^2 C_\mu - \pro_\mu \pro_\nu C_\nu),
\label{qc}
\eea
which also has two components, the component
moving along the knot, and the one moving around
the knot tube.
Now it must be clear that the supercurrent
moving along the knot generates an angular momentum
around the $z$-axis which provides the centrifugal force
preventing the vortex ring to collapse.
Put it differently, the supercurrent generates the $m$ unit of
the magnetic flux trapped in the knot disk
which can not be squeezed out. And this flux provides
a stablizing repulsive force which prevent the collapse of the knot.
This is how the knot acquires the dynamical stability.
It is this remarkable interplay between topology
and dynamics which assures the existence of the stable knot
in Skyrme theory. The nontrivial topology
of the magnetic flux rings which provides
the topological stability now expresses itself
in the form of the supercurrent and angular momentum
which provides the dynamical stability of the knot.

The above analysis also makes it clear that alternatively
the knot can also be viewed as a two quantized vorticity rings
linked together in a two component superfluid, whose
linking number gives the knot quantum number.

The energy of the Faddeev-Niemi knot is known to have
the following bound \cite{ussr},
\bea
c_1~Q^{3/4} \leq E_Q \leq c_2~Q^{3/4},
\eea
which implies that the energy is proportional to $Q^{3/4}$.
Numerically this has been confirmed up to $Q=8$ \cite{batt}
\bea
E_Q \simeq 252~Q^{3/4} \sqrt \alpha \mu
\eea
This means that knot with large $Q$ can not decay to
the knots with smaller $Q$. 

We close with the following remarks. \\
1. In this paper we have clarified the physical meaning
of topological objects in Skyrme theory.
In particular, we have shown that
the Faddeev-Niemi knot in Skyrme theory is
nothing but the chromomagnetic vortex ring made of
a monopole-antimonopole flux, twisted and
connected together. This picture allows us to interprete
the knot as two quantized flux rings linked together,
whose knot quantum number is given by the linking number
of the rings. This interpretation
follows from the fact that the Lagrangian (\ref{sflag})
can be viewed as a massive Yang-Mills Lagrangian, which
emphasizes the deep connection between the Skyrme theory
and QCD \cite{cho02}.  \\
2. Our anslysis tells that the Lagrangian (\ref{sflag})
could also be understood to describe a theory of two-component
superfluid. This implies that it could
play an important role in condensed matter physics,
which puts the Skyrme theory
in a totally new perspective.
The Skyrme theory has always
been associated to nuclear and/or high energy physics.
But now it becomes clear
that the theory could also describe interesting low energy physics
in a completely different environment,
in two-component condensed matters
\cite{cho1,cho2}. This is really remarkable. \\
3. In our analysis we have outlined
how one can actually construct the
Faddeev-Niemi knot (or a similar one) in laboratories.
So the challenge now is to verify the existence of the topological
knot experimentally. Constructing the knot may be a tricky task 
at present moment, but might have already been done 
in two-component Bose-Einstein condensates \cite{exp2,exp3}.
We predict that similar knots could be constructed 
in laboratories in the near future.

 A detailed discussion on the subject will be published 
elsewhere \cite{sky3}.

{\bf ACKNOWLEDGEMENT}

~~~The author thanks G. Sterman for the kind hospitality during
his visit to Institute for Theoretical Physics. 
The work is supported in part by the BR 
Program and ABRL Program of Korea Science and Enginnering 
Foundation (Grant R02-2003-000-10043-0
and Grant R14-2003-012-01002-0), and by the BK21 Project of 
the Ministry of Education.

\end{document}